\documentstyle[epsf]{aipproc}

\newcommand{\be}{\begin{equation}}
\newcommand{\ee}{\end{equation}}
 
\newcommand{\bea}{\begin{eqnarray}}
\newcommand{\eea}{\end{eqnarray}}
\newcommand{\ba}{\begin{array}}
\newcommand{\ea}{\end{array}}
\newcommand{\beqa}{\begin{eqnarray}}
\newcommand{\eeqa}{\end{eqnarray}}
 
\newcommand{\NP}[1]{Nucl. Phys.\ {\bf #1}}
\newcommand{\PL}[1]{Phys. Lett.\ {\bf #1}}

\newcommand{\PRD}[1]{Phys. Rev.\ {\bf #1}}

\newcommand{\PRL}[1]{Phys. Rev. Lett.\ {\bf #1}}

\newcommand{\ZP}[1]{Z. Phys.\ {\bf #1}}

\newcommand{\Tr}{{\rm Tr}}

\newcommand{\ssu}{$SU(2)_L\times SU(2)_R\times U(1)_{B-L}\,$}

\newcommand{\matr}{\left( \begin{array}}
\newcommand{\ematr}{\end{array} \right)}

\newcommand{\lsim}
{{\;\raise0.3ex\hbox{$<$\kern-0.75em\raise-1.1ex\hbox{$\sim$}}\;}}
\newcommand{\gsim}
{{\;\raise0.3ex\hbox{$>$\kern-0.75em\raise-1.1ex\hbox{$\sim$}}\;}}
 
\begin{document}
\title{Constraining the scales of \\ supersymmetric left-right 
models\footnote{Talk given in Beyond the Standard Model V in Balholm,
Norway.  Presented by K. Huitu.}}

\author{
{K. Huitu$^*$, P.N. Pandita$^{*,\dagger}$
and K. Puolam\"aki$^*$}}
\address{$^*$Helsinki Institute of Physics, 
FIN-00014 University of Helsinki,
Finland
\\$^\dagger$Department of Physics, North Eastern Hill University,
Shillong 793022, India\thanks{Permanent address} }

\maketitle

\begin{abstract}
We'll review our study of the constraints on the scales in the
supersymmetric left-right model (SUSYLR).
The conservation of color and electromagnetism in the ground state
of the theory implies a relation between right-handed gauge boson 
mass and soft squark mass.
Furthermore, in general for heavy $W_R$, $\tan\alpha$ is larger 
than one, and the
right-handed sneutrino VEV, responsible for spontaneous $R$-parity
breaking, is at most of the order $M_{SUSY}/h_{\Delta_R}$, where
$M_{SUSY}$ is supersymmetry breaking scale and $h_{\Delta_R}$ is the
Yukawa coupling in Majorana mass term for right-handed neutrinos.
\end{abstract}

\section*{Introduction}

The conservation of baryon (B) and lepton (L) numbers, which is automatic
in the Standard Model, is not apparent in low energy supersymmetric models.
In MSSM, baryon and lepton
number violation can occur at tree level with fast proton decay unless
the corresponding couplings are very small.
The most common way to eliminate these tree level B and L violating terms is 
to impose so-called R-parity, $R_p=(-1)^{3(B-L)+2S}$, where $S$ is the spin
of the particle.
However, R-parity conservation is not required for the internal 
consistency of the minimal supersymmetric standard model.

It would be more appealing to have a supersymmetric theory,
where R-parity is related to a gauge symmetry, and its conservation
is automatic because of the invariance of the underlying theory under an 
extended gauge symmetry.
Indeed $R_p$ conservation follows automatically in certain theories with
gauged (B-L).
It has been noted by several authors \cite{Rauto1,Rauto2} that if the gauge 
symmetry 
of MSSM is extended to $SU(2)_L\times U(1)_{I_{3R}}\times U(1)_{B-L}$ or
$SU(2)_L\times SU(2)_R\times U(1)_{B-L}$, the theory becomes
automatically R-parity conserving.
Such a left-right supersymmetric theory (SUSYLR) solves the problems of
explicit B and L violation of MSSM and has received much attention
recently $\left[\right.$3-7$\left.\right]$.

It has been found in a wide class of such models \cite{km} that R-parity must 
be 
spontaneously broken \cite{am} because of the form of the scalar potential.
Thus in the model we have several different scales, namely
the $SU(2)_R$ breaking scale, R-parity breaking scale, SUSY breaking scale
and finally the weak scale.
For phenomenological reasons, it would be desirable to relate these scales
with each other.

It has been shown \cite{km2} in the minimal SUSYLR model
that the mass ($m_{W_R}$) of the right-handed gauge boson $W_R$ has an upper
limit related to the SUSY breaking scale, {\it i.e.}, 
$m_{W_R}\leq g M_{SUSY}/h_{\Delta_R}$, where $g$ is the weak gauge coupling and
$h_{\Delta_R}$ is the Yukawa coupling of the right-handed neutrinos 
with the triplet Higgs fields.
Here we review  some further constraints on the scales  \cite{hpp1}
following from the conservation of electric charge and color 
by the ground state of the theory.

\section*{Supersymmetric left-right model}

The quark and lepton superfields are denoted by
$Q(2,1,1/3)$; $Q^c(1,2,-1/3)$; $L(2,1,-1)$; $L^c(1,2,1)$,
and the Higgs superfields by $\Delta_L(3,1,-2)$; $\Delta_R(1,3,-2)$;
$\delta_L(3,1,2)$; $\delta_R(1,3,2)$; $\Phi (2,2,0)$; $\chi (2,2,0)$.
The numbers in the parantheses denote the representation content of the
fields under the gauge group \ssu.
The model is described by the superpotential 

\bea
W&=& h_{\phi Q}Q^T i\tau_2 \Phi Q^c +h_{\chi Q}Q^T i\tau_2 \chi Q^c +
h_{\phi L}L^T i\tau_2 \Phi L^c 
+ h_{\chi L}L^T i\tau_2 \chi L^c \nonumber\\
&&+h_{\delta_L} L^T i\tau_2 \delta_L L +
h_{\Delta_R} L^{cT} i\tau_2 \Delta_R L^c+
\mu_1 \Tr (i\tau_2\Phi^T i\tau_2 \chi) +
\mu_1' \Tr (i\tau_2\Phi^T i\tau_2 \Phi) \nonumber\\
&&+ \mu_1'' \Tr (i\tau_2\chi^T i\tau_2 \chi) 
+\Tr (\mu_{2L}\Delta_L \delta_L +
\mu_{2R}\Delta_R\delta_R).
\eea

The vacuum expectation values of various scalar fields
which preserve electric charge can be written as
\bea
&&\langle \Phi\rangle  = \matr {cc} \kappa_1&0\\0&e^{i\varphi_1}\kappa '_1 
\ematr 
,\;\;
\langle \chi\rangle  = \matr {cc} e^{i\varphi_2}\kappa '_2&0\\0&\kappa_2 
\ematr ,
\nonumber\\
&& \langle \Delta_L\rangle = \matr {cc} 0&v_{\Delta_L}\\0&0\ematr ,\;
\langle \delta_L\rangle = \matr {cc} 0&0\\v_{\delta_L}&0\ematr , \nonumber\\
&&\; \langle \Delta_R\rangle = \matr {cc} 0&v_{\Delta_R}\\0&0\ematr ,\;
\langle \delta_R\rangle = \matr {cc} 0&0\\v_{\delta_R}&0\ematr ,\nonumber\\
&&\langle L\rangle =\matr {c} \sigma_L\\0\ematr,\;
\langle L^c\rangle =\matr {c} 0\\\sigma_R\ematr .
\eea
The phases $\varphi_1$ and $\varphi_2$ are ignored
in the following, although this does not affect the final conclusion.
Due to the tiny mixing between the charged gauge bosons, $\kappa '_1$
and $\kappa '_2$ are taken to be much smaller than $\kappa_1$ and $\kappa_2$.
Furthermore, since the electroweak $\rho$-parameter is close to unity,
$\rho=1.0002\pm0.0013\pm0.0018 $ \cite{pdg}, the triplet vacuum expectation
values $\langle\Delta_L\rangle $ and $\langle\delta_L\rangle $ must be small.
For definiteness, we shall take 
$v_{\Delta_R}\sim v_{\delta_R}\sim v_R$, the generic scale of the
right-handed symmetry breaking.

It has been shown on general grounds \cite{km} and explicitly
\cite{hm} that due to the $U(1)_{em}$ symmetry of the ground state of the
model, the spontaneous breakdown of R-parity is inevitable 
in this class of models.
Thus at least one of the sneutrinos has VEV in the minimum of the
potential.
We shall assume that $\sigma_L$ and $\sigma_R$
are non-zero.
In electric charge preserving ground state
$\sigma_R$ is necessarily at least of the
order of the typical
SUSY breaking scale $M_{SUSY}$  or the right-handed breaking scale
$v_R$,
whichever is lower \cite{km2}.

\section*{Constraints from squark masses}

It is straightforward to find the squark masses from the scalar potential 
of the model \cite{hpp1}.
We'll consider the up- and down-squark mass matrices for the 
lightest generation (ignoring the intergenerational mixing) which gives
the tightest constraint in our case.
The part of the potential containing the squark mass terms can be
written as
\be
V_{squark}=\matr {cc}   U_L^* &   U_R^* \ematr \tilde M_U
\matr {c}   U_L \\   U_R \ematr +
\matr {cc}  D_L^* &   D_R^* \ematr \tilde M_D
\matr {c}   D_L \\   D_R \ematr .
\ee
\noindent 
The diagonal mass matrix elements for the up-type squarks
($g_L,\; g_R,\; g_{B-L}$ are the gauge couplings and  $\tilde m_Q^2$
$\tilde m_{Q^c}^2$ are the soft squark masses) are given by
\bea
(\tilde M_U)_{  U_L^*   U_L} &=&  \tilde m_Q^2+
m_u^2 +
\frac 14 g_L^2(\omega^2_\kappa -2\omega^2_L ) +
\frac 16 g_{B-L}^2(\omega^2_L -\omega^2_R ) ,
\nonumber\\
(\tilde M_U)_{  U_R^*   U_R} &=&\tilde m_{Q^c}^2+
m_u^2 +
\frac 14 g_R^2(\omega^2_\kappa -2\omega^2_R ) +
\frac 16 g_{B-L}^2(\omega^2_R -\omega^2_L ) 
,
\eea
\noindent
and for down-type squarks
\bea
(\tilde M_D)_{  D_L^*   D_L} &=&
 \tilde m_Q^2+
m_d^2
-\frac 14 g_L^2(\omega^2_\kappa -2\omega^2_L ) +
\frac 16 g_{B-L}^2(\omega^2_L -\omega^2_R ) 
,\nonumber\\
(\tilde M_D)_{ D_R^*  D_R  } &=&\tilde m_{Q^c}^2+
m_d^2
-\frac 14 g_R^2(\omega^2_\kappa -2\omega^2_R ) +
\frac 16 g_{B-L}^2(\omega^2_R -\omega^2_L ) ,
\eea
\noindent
where
\be
m_u=h_{\phi Q}\kappa '_1+h_{\chi Q}\kappa_2,\;
m_d=h_{\phi Q}\kappa_1+h_{\chi Q}\kappa '_2.
\ee
\noindent
and
\bea
\omega^2_L=v_{\delta_L}^2-v_{\Delta_L}^2-\frac 12\sigma_L^2,\;
\omega^2_R=v_{\Delta_R}^2-v_{\delta_R}^2-\frac 12\sigma_R^2,\;
\omega^2_\kappa = \kappa_1^2+{\kappa '_2}^2-\kappa_2^2-{\kappa '_1}^2.
\eea
In order not to break electromagnetism or color, none of the physical
squared masses of squarks can be negative.
Thus necessarily all the diagonal elements of the squark mass matrices
are non-negative.
Next we define an angle $\alpha $ by
\bea\tan^2\alpha =(v_{\delta_R}^2 +\frac 12 \sigma_R^2)/ v_{\Delta_R}^2\eea
and write
$\tilde m_Q^2=\tilde m_{Q^c}^2\equiv\tilde m^2$.
We then recall that the right-handed
gauge boson mass is given by (ignoring weak scale effects) \cite{hm}
\be
m_{W_R}^2=g_R^2 (v_{\Delta_R}^2+v_{\delta_R}^2 +\frac 12\sigma_R^2 )
=g_R^2 v_{\Delta_R}^2 (1+\tan^{2}\alpha ).
\label{mwr}
\ee
Then combining the diagonal elements of the mass matrices $\tilde M_U $ and
$\tilde M_D$, and 
ignoring terms of the order of the weak scale or smaller, it follows that
\be
m_{W_R}^2|\cos 2\alpha|\leq 4\tilde m^2.
\label{result}
\ee
If the $W_R$ boson is lighter than twice the soft squark mass $\tilde m$,
Eq.(\ref{result}) 
is fulfilled for any $\tan\alpha $.
If $\tilde m\sim M_{SUSY}\sim 1$ TeV as is commonly assumed, 
$m_{W_R}$ cannot be 
much less, since experimentally  $m_{W_R}>420$ GeV \cite{cdf}.
On the other hand,
if $m_{W_R}$ is much larger than $2\tilde m$, $\tan\alpha$ has to be close
to one, e.g. for $m_{W_R}=10$ TeV and $\tilde m$=1 TeV, one would need
$0.96\leq\tan\alpha\leq 1.04$.
It is interesting to note in this context that 
the vanishing of $D$-terms implies $\tan\alpha=1$.
To translate the limit for $\tan\alpha $ to  an upper bound for the 
VEV $\langle \Delta_R^0\rangle$, one needs a lower limit for
$g_R$.
This was found in \cite{cl} from 
$\sin^2\theta_W=e^2/g_L^2=0.23$, namely $g_R \geq 0.55\; g_L$.
Consequently 
$v_{\Delta_R}\leq m_{W_R} |\cos\,\alpha |/(0.55 \, g_L)$, e.g.
in our example $v_{\Delta_R}\lsim 20$ TeV.

\section*{Constraints from doubly charged higgs masses}

In the case of large $m_{W_R}$,
we note that if the right-handed scale and R-parity breaking scale differ
from each other, one has  $v_{\Delta_R},\; v_{\delta_R}
>\sigma_R$, since $\tan\alpha\sim 1$.
We recall then the doubly charged Higgs mass matrix \cite{hm} 
 given by
(ignoring terms suppressed by $\sigma_R/v_{\Delta_R}$ or
$\sigma_R/v_{\delta_R}$) 
\bea
&&M_{\Delta^{++}\delta^{++}}^2=
\matr{cc} m_{\Delta\delta}^2\frac{v_\delta}{v_\Delta}
-4h_\Delta^2\sigma_R^2-2g_R^2\omega_R^2
&-m_{\Delta\delta}^2 \\ -m_{\Delta\delta}^2 &
m_{\Delta\delta}^2\frac{v_\Delta}{v_\delta}+
2g_R^2\omega_R^2 \ematr,
\eea
where $ m_{\Delta\delta}$ is the soft parameter mixing right-handed
Higgs triplets.
The two eigenvalues of the mass matrix need to be real and non-negative
in order not to break $U(1)_{em}$.
This leads to two conditions:
\bea
&& h_{\Delta_R}^2\sigma_R^2 \leq \frac 14
m_{\Delta\delta}^2\left( \frac {v_{\delta_R}}{v_{\Delta_R}} +
 \frac {v_{\Delta_R}}{v_{\delta_R}}\right)
\label{cond1}
\eea
and
\bea
 -2m_{\Delta\delta}^2g_R^2\omega_R^2
\left( \frac {v_{\delta_R}}{v_{\Delta_R}} -
 \frac {v_{\Delta_R}}{v_{\delta_R}}\right)+4 g_R^4\omega_R^4 
+4 m_{\Delta\delta}^2h_{\Delta_R}^2\sigma_R^2\frac {v_{\Delta_R}}{v_{\delta_R}}
+ 8 h_{\Delta_R}^2\sigma_R^2 g_R^2\omega_R^2 
\leq 0
\label{cond2}
\eea

{}From (\ref{cond1}) we see that $h_{\Delta_R}\sigma_R$ can be at most of the
order of $m_{\Delta\delta}\sim M_{SUSY}$.
To fulfill the inequality  (\ref{cond2}) we can consider two
cases: $v_{\Delta_R}<v_{\delta_R}$ and  $v_{\delta_R}<v_{\Delta_R}$.
In both cases one must have $\omega_R^2\leq 0$ or
equivalently $\tan\alpha \geq 1$.
The equality can hold for $\sigma_R=0$ and $v_{\delta_R}=v_{\Delta_R}$.

\section*{Summary}

In this talk we discussed the constraints on scales in
SUSY left-right model.
The $W_R$ mass and the soft squark mass are related
by (\ref{result}), which implies that either
the scale of the right-handed gauge symmetry breaking must
be close to the SUSY breaking scale,
or $\tan\alpha \sim 1$ corresponding to vanishing $D$-terms.
In general, for large $m_{W_R}$, 
the right-handed sneutrino VEV is constrained to be at most of the
order $M_{SUSY}/h_{\Delta_R}$, and $\tan\alpha $ is larger than one.

\vspace*{0.2cm}
\noindent
{\bf Acknowledgements}

One of us (PNP) would like to thank the Helsinki Institute of Physics for 
hospitality while this work was completed.
The work of PNP is supported by the Department of Atomic Energy Project
No.37/14/95-R \& D-II/663.

\end{document}